\documentstyle[sprocl,epsfig]{article}

\input{psfig}
\def\Journal#1#2#3#4{{#1}{\bf #2}, #3 (#4)}

\def\NPB{{\em Nucl. Phys.} B}
\def\NPBP{{\em Nucl. Phys.} B {\em (Proc. Suppl.)}}
\def\PLB{{\em Phys. Lett.} B}
\def\PRL{\em Phys. Rev. Lett. }
\def\PRD{{\em Phys. Rev.} D}

\def\ZPB{{\em Z. Phys.} B}
\def\RMP{\em Rev. Mod. Phys. }
\def\AP{\em Ann. Phys. }
\def\JCP{\em J. Chem. Phys. } 
\newcommand{\Z}{{\sf Z \!\!\! Z}}
\newcommand{\R}{{\sf I \!\! R}}

\newcommand{\1}{{\sf 1 \!\! 1}}

\begin{document}

\title{THE CENTER SYMMETRY AND ITS SPONTANEOUS BREAKDOWN AT HIGH TEMPERATURES}

\author{KIERAN HOLLAND}

\address{Institute for Theoretical Physics, University of Bern, CH-3012 Bern,
Switzerland}
\author{UWE-JENS WIESE}

\address{Center for Theoretical Physics, \\
Laboratory for Nuclear Science and Department of Physics, \\ 
Massachusetts Institute of Technology, Cambridge, Massachusetts 02139, USA}

\maketitle

\abstracts{The Euclidean action of non-Abelian gauge theories with adjoint 
dynamical charges (gluons or gluinos) at non-zero temperature $T$ is invariant 
against topologically non-trivial gauge transformations in the $\Z(N)_c$ center
of the $SU(N)$ gauge group. The Polyakov loop measures the free energy of 
fundamental static charges (infinitely heavy test quarks) and is an order 
parameter for the spontaneous breakdown of the center symmetry. In $SU(N)$ 
Yang-Mills theory the $\Z(N)_c$ symmetry is unbroken in the low-temperature 
confined phase and spontaneously broken in the high-temperature deconfined 
phase. In 4-dimensional $SU(2)$ Yang-Mills theory the deconfinement phase 
transition is of second order and is in the universality class of the 
3-dimensional Ising model. In the $SU(3)$ theory, on the other hand, the 
transition is first order and its bulk physics is not universal. When a 
chemical potential $\mu$ is used to generate a non-zero baryon density of test 
quarks, the first order deconfinement transition line extends into the 
$(\mu,T)$-plane. It terminates at a critical endpoint which also is in the 
universality class of the 3-dimensional Ising model. At a first order phase 
transition the confined and deconfined phases coexist and are separated by 
confined-deconfined interfaces. Similarly, the three distinct high-temperature 
phases of $SU(3)$ Yang-Mills theory are separated by deconfined-deconfined 
domain walls. As one approaches the deconfinement phase transition from the 
high-temperature side, a deconfined-deconfined domain wall splits into a pair 
of confined-deconfined interfaces and becomes completely wet by the confined 
phase. Complete wetting is a universal interface phenomenon that arises despite
the fact that the bulk physics is non-universal. In supersymmetric $SU(3)$ 
Yang-Mills theory, a $\Z(3)_\chi$ chiral symmetry is spontaneously broken in 
the confined phase and restored in the deconfined phase. As one approaches the 
deconfinement phase transition from the low-temperature side, a 
confined-confined domain wall splits into a pair of confined-deconfined 
interfaces and thus becomes completely wet by the deconfined phase. This allows
a confining string to end on a confined-confined domain wall as first suggested
by Witten based on M-theory. Deconfined gluons and static test quarks are 
sensitive to spatial boundary conditions. For example, on a periodic torus the 
Gauss law forbids the existence of a single static quark. On the other hand, on
a C-periodic torus (which is periodic up to charge conjugation) a single static
quark can exist. As a paradoxical consequence of the presence of 
deconfined-deconfined domain walls, in very long C-periodic cylinders quarks 
are ``confined'' even in the deconfined phase.}
  
\section{Introduction and Summary}

Understanding the non-perturbative dynamics of quarks and gluons is a highly
non-trivial task. Since at low temperatures these particles are confined into 
hadrons, the quark and gluon fields that appear in the QCD Lagrangian do not 
provide direct insight into the non-perturbative physics of confinement. In the
real world with light quarks confinement is accompanied by spontaneous chiral
symmetry breaking and the presence of light pseudo-Goldstone pions. The pion 
fields that appear in the effective chiral Lagrangian do provide direct 
physical insight into the low-energy physics of QCD, but they do not shed light
on the dynamical mechanism that is responsible for chiral symmetry breaking or 
confinement itself. 

In order to gain some insight into the physics of confinement, it is very
interesting to study QCD in extreme conditions of high temperature and large 
baryon chemical potential. Due to asymptotic freedom, confinement is then lost 
and we can hope to learn more about confinement by understanding how it turns 
into deconfinement. For the theory with light quarks this question is still
very difficult to address from first principles, although a lot of interesting 
results have been obtained using lattice field theory as well as 
phenomenological models.

Here we decide to study a hypothetical world without dynamical quarks. Then the
only dynamical color charge carriers are the gluons. Quarks appear only as
infinitely heavy test charges that allow us to probe the gluon dynamics. In the
resulting Yang-Mills theory confined and deconfined phases are easier to 
understand than in full QCD because they are characterized by distinct
symmetry properties. As first pointed out by 't Hooft \cite{tHo78}, the 
symmetry that distinguishes confinement from deconfinement is the $\Z(N)_c$ 
center symmetry of the $SU(N)$ gauge group. 

The selection of topics covered in this article is strongly biased by the 
authors' personal interests. Still, we hope that it will give a useful overview
of the many interesting phenomena related to the center symmetry and its 
spontaneous breakdown. There is an extensive literature on the subject and we
will not be able to discuss all important contributions. We like to apologize 
to those whose work will not be mentioned here.

\subsection{Static Quarks and Dynamical Gluons}

Gluons transform in the adjoint representation of $SU(N)$ and are $\Z(N)_c$ 
neutral. A single test quark transforms in the fundamental representation and 
carries a unit of $\Z(N)_c$ charge, while $N$ test quarks (forming a ``test 
baryon'') are together $\Z(N)_c$ neutral. In Yang-Mills theory a confined phase
is simply characterized by the fact that a non-zero $\Z(N)_c$ charge (for
example, a single test quark) costs an infinite amount of free energy. In a 
deconfined phase, on the other hand, the free energy of a single static quark 
is finite. In full QCD this simple confinement criterion fails because in the 
presence of dynamical quarks the free energy of a single static test quark is 
always finite. This is because the color flux string emanating from the static 
quark can end in a dynamical anti-quark. The static-quark-dynamical-anti-quark 
meson has a finite free energy. In Yang-Mills theory, on the other hand, there 
are no dynamical quarks or anti-quarks and the string emanating from a test 
quark is absolutely unbreakable. Since it carries the $\Z(N)_c$ flux of the 
test quark, the string cannot end in a dynamical gluon which is $\Z(N)_c$ 
neutral. Instead it must go to infinity. Since the confining string has a 
non-zero tension (free energy per unit length) the static quark with the 
infinite flux string emanating from it costs an infinite amount of free energy.

Polyakov \cite{Pol78} and Susskind \cite{Sus79} were first to point out that 
the operator that describes a static quark is a Wilson loop closed around the 
periodic Euclidean time direction --- the so-called Polyakov loop $\Phi$. The 
free energy $F$ of the test quark is determined by the thermal expectation 
value of the Polyakov loop as $\langle \Phi \rangle = \exp(- \beta F)$. Hence, 
in a confined phase in which $F = \infty$ one has $\langle \Phi \rangle = 0$. 
This can be understood as follows. At low temperatures $T$ the extent 
$\beta = 1/T$ of the Euclidean time direction is large. Then the Polyakov loop 
extends through many space-time regions that are essentially uncorrelated in 
color space. As a consequence, the Polyakov loop behaves essentially as a 
random variable and its thermal expectation value averages to zero. In the 
high-temperature deconfined phase, on the other hand, $F$ is finite and hence 
$\langle \Phi \rangle \neq 0$. In that case the Euclidean time direction is
short and the Polyakov loop extends only through a small space-time region 
that is highly correlated in color space. Then the Polyakov loop orders and
picks up a non-zero expectation value.

Under a twisted gauge transformation that is periodic up to an element $z$ of 
the center $\Z(N)_c$, the Polyakov loop changes by a factor $z$, while the 
action remains invariant. Consequently, the Polyakov loop is an order parameter
for center symmetry breaking. A non-zero value of $\langle \Phi \rangle$ 
implies that the center symmetry is spontaneously broken. This is the case in 
the high-temperature deconfined phase. In the confined phase, on the other 
hand, the center symmetry is unbroken. This result is counter-intuitive, 
because one might expect spontaneous symmetry breaking to occur only at low 
temperatures. However, it should also not be too surprising because, for 
example, even the Ising model displays spontaneous symmetry breaking at high 
temperature when it is formulated in terms of dual variables. It should be 
noted that the center symmetry is absent in full QCD with dynamical quarks 
because their presence leads to explicit $\Z(3)_c$ breaking.

\subsection{Phase Transitions and Interface Dynamics}

As the order parameter that distinguishes confinement from deconfinement, the
Polyakov loop plays a central role in understanding the physics of the 
high-temperature phase transition. In fact, just as the pion field of the 
effective chiral Lagrangian provides direct insight into the low-energy physics
of QCD, an effective action for the Polyakov loop provides direct insight into 
the high-temperature physics close to the confinement-deconfinement phase 
transition. In the case of $SU(2)$ Yang-Mills theory, the Polyakov loop is
represented by a real-valued scalar field in three dimensions with a $Z(2)_c$
symmetry. Hence, the effective theory for the Polyakov loop is a 3-d $\Phi^4$
theory. This theory is in the universality class of the 3-d Ising model. It was
first noted by McLerran and Svetitsky \cite{McL81a} and by Svetitsky and Yaffe 
\cite{Sve82} that the second order phase transition in the 3-d Ising model has 
the same universal features as the deconfinement phase transition in the 
$SU(2)$ Yang-Mills theory. The first numerical lattice simulations indicating a
second order $SU(2)$ deconfinement phase transition were performed in 
\cite{McL81b}. A detailed analysis of the critical exponents \cite{Eng90} shows
that they are indeed the same as in the 3-d Ising model. The lattice 
calculations of the value of the critical temperature have been extrapolated 
reliably to the continuum limit \cite{Fin93}.

In $SU(3)$ Yang-Mills theory the Polyakov loop is described by a complex-valued
field $\Phi$ in three dimensions. The effective theory is again a 3-d $\Phi^4$ 
theory but now with a $\Z(3)_c$ symmetry. A lattice model with the same 
symmetries is the 3-d 3-state Potts model \cite{Wu82} which is known to have a 
first order phase transition \cite{Kna79}. In $SU(3)$ Yang-Mills theory the 
deconfinement phase transition is therefore also expected to be first order 
\cite{Yaf82} so that the bulk physics is not universal. Lattice gauge theory 
has indeed provided detailed numerical evidence for a first order transition in
$SU(3)$ Yang-Mills theory \cite{Cel83}. At a first order phase transition both 
phases coexist with each other and are separated by interfaces. For example, at
the deconfinement phase transition the low-temperature confined phase and the 
high-temperature deconfined phase are separated by confined-deconfined 
interfaces. The simplest mechanism for the dynamics of a first order phase 
transition is bubble nucleation. For example, in boiling water bubbles of the 
gas phase form and grow in the liquid phase. Similarly, when one heats up a 
system of glueballs (confined gluons) beyond the temperature of the 
deconfinement phase transition, one expects the formation of bubbles of gluon 
plasma (deconfined gluons) which grow and ultimately fill the entire volume 
with deconfined phase. Due to the $\Z(3)_c$ symmetry, there are in fact three 
distinct deconfined phases. Hence, bubbles of different deconfined phases may 
nucleate independently in different parts of the volume. When these bubbles 
expand and collide they cannot coalesce and form a single bubble because they 
contain distinct types of deconfined phase. Consequently, when the phase 
transition is completed, we are left with different deconfined regions 
separated by deconfined-deconfined domain walls. These domain walls can exist 
at all temperatures above the phase transition point, while confined-deconfined
interfaces are stable only at the phase transition. Deconfined-deconfined 
domain walls have a non-zero tension (free energy per unit area). Due to 
asymptotic freedom the domain wall tension can be calculated analytically at 
high temperatures using semi-classical methods \cite{Bha91}. The 
confined-deconfined interface tension, on the other hand, is accessible only to
lattice simulations \cite{Hua90} because the phase transition itself occurs at 
a temperature at which analytic methods based on asymptotic freedom are not 
applicable. It should be noted that in full QCD deconfined domains are unstable
or at most meta-stable, because dynamical quarks break the center symmetry 
explicitly. The fate of a meta-stable deconfined domain in the early universe 
has been discussed \cite{Ign92}. Via a tunneling process it turns into the 
stable deconfined phase at temperatures around 10 TeV, i.e. much before it 
reaches the deconfinement phase transition at temperatures around 100 MeV.

When one cools the gluon plasma down below the phase transition temperature, 
one expects bubbles of the confined phase to nucleate in the bulk deconfined 
phase. These bubbles grow and ultimately fill the entire volume with confined
(glueball) phase. Interestingly, in the presence of deconfined-deconfined
domain walls there is a competing mechanism for generating the confined phase.
Rather than nucleating in the deconfined bulk (which costs a large amount of 
confined-deconfined interface tension), it is energetically favorable for the
confined phase to appear at a deconfined-deconfined domain wall, thus making
use of the free energy provided by the domain wall tension \cite{Fre89}. In 
fact, the confined phase appears as a thin film that completely covers the 
domain wall. Hence, the deconfined-deconfined domain wall splits into a pair of
confined-deconfined interfaces. Such phenomena are well known in condensed
matter physics under the name of complete wetting. For example, in the human
eye a film of liquid of tears completely wets the eye and thus separates it 
from the surrounding air. The alternative to complete wetting is incomplete
wetting. Then lens shaped droplets of one phase appear at the domain wall
separating two other phases. Interestingly, complete wetting is a universal
phenomenon that occurs despite the fact that the phase transition is first
order and the bulk physics is non-universal. In fact, complete wetting cannot
occur at a second order bulk phase transition because it requires interfaces 
and thus phase coexistence. Complete wetting is characterized by a set of 
universal critical exponents which can be extracted from the effective 3-d 
$\Z(3)_c$ symmetric $\Phi^4$ theory for the Polyakov loop \cite{Tra92}.

The $\Z(3)_c$ center symmetry also plays an important role in the 
high-temperature physics of supersymmetric Yang-Mills theory. The supersymmetry
partners of the gluons are adjoint Weyl fermions --- the gluinos. Like gluons, 
gluinos are $\Z(3)_c$ neutral and their confinement or deconfinement can again 
be characterized by the value of the Polyakov loop. In particular, the 
$\Z(3)_c$ flux of an external fundamental test quark cannot end at a dynamical 
adjoint gluino and must hence go to infinity. In the confined phase this still 
costs an infinite amount of free energy. Besides the $\Z(3)_c$ symmetry, 
supersymmetric Yang-Mills theory has a $\Z(3)_\chi$ chiral symmetry. This 
symmetry is a remnant of the $U(1)_R$ symmetry that is explicitly broken down 
to $\Z(3)_\chi$ by an anomaly. In the low-temperature confined phase the 
$\Z(3)_\chi$ symmetry is spontaneously broken by the formation of a gluino 
condensate, while the $\Z(3)_c$ symmetry remains unbroken. At high temperatures
one expects chiral symmetry to be restored and the $\Z(3)_c$ symmetry to be 
spontaneously broken due to deconfinement.

Due to the spontaneous breakdown of the $\Z(3)_\chi$ chiral symmetry there are
three distinct confined phases in supersymmetric $SU(3)$ Yang-Mills theory that
are distinguished by the value of the gluino condensate. Below the 
deconfinement phase transition these phases coexist and are separated by
confined-confined domain walls. When such a domain wall is heated up to the
phase transition temperature, one of the three high-temperature deconfined
phases forms a complete wetting layer and the confined-confined domain wall
splits into a pair of confined-deconfined interfaces \cite{Cam98}. The fact 
that the wetting layer consists of deconfined phase has interesting 
consequences. It allows a color flux string emanating from a static test quark 
to end on the wall because inside the deconfined wetting layer the flux spreads
out and no longer costs infinite energy when transported to infinity. The 
phenomenon of strings ending on walls in supersymmetric Yang-Mills theory was 
first predicted by Witten in the context of M-theory \cite{Wit97}. The color 
flux string then appears as a fundamental string and the confined-confined 
domain wall manifests itself as a D-brane --- a natural place for strings to 
end. Complete wetting provides a field theoretical explanation of this 
phenomenon.

\subsection{The Center --- A Subtle Symmetry}

It should be noted that --- unlike other symmetries --- the center symmetry is 
not a symmetry of the Hamiltonian which describes the time evolution of the 
system. Instead, it is a symmetry of the transfer matrix which describes the 
spatial evolution of the system in thermal equilibrium. This subtle difference
is at the origin of a controversy about the physical reality of distinct
high-temperature phases and of deconfined-deconfined domain walls. It has been 
argued that these domain walls are just Euclidean field configurations (like 
instantons) which play no role in Minkowski space-time \cite{Bel92}. For 
example, in contrast to other topological excitations like monopoles or cosmic 
strings, one cannot expect a deconfined-deconfined domain wall to appear as an 
eigenstate of the theory's Hamiltonian. This should not be too surprising 
because the deconfined phases themselves are thermal mixtures of pure states 
weighted by their Boltzmann factors. Of course, the entire spectrum of the 
Hamiltonian is completely independent of temperature. The spectrum of the 
transfer matrix, on the other hand, is very sensitive to the temperature which 
controls the size of the Euclidean time dimension. Despite the fact that 
deconfined-deconfined domain walls are not energy eigenstates, their presence 
in the thermally equilibrated gluon plasma has observable consequences --- at 
least in Gedanken experiments.\footnote{In the real world the presence of 
dynamical quarks leads to an explicit $\Z(3)_c$ breaking which eliminates two 
of the three deconfined phases as well as the deconfined-deconfined domain 
walls.} For example, the free energy of a static test quark in a large box 
filled with gluon plasma depends crucially on the shape of the box. This effect
is due to the presence of distinct high-temperature phases and the 
deconfined-deconfined domain walls that separate them. 

To study the free energy of a single static quark in a finite box one must 
choose appropriate boundary conditions. Standard periodic boundary conditions
are inappropriate because --- as a consequence of the $\Z(3)_c$ Gauss law ---
a single quark cannot even exist on a torus \cite{Hil83}. In a periodic volume 
the color flux emanating from a quark cannot go to infinity and must thus end 
in an anti-quark. The system as a whole is thus $\Z(3)_c$ neutral. In a finite
periodic volume the expectation value of the Polyakov loop always vanishes even
in the deconfined phase. As a consequence of the $\Z(3)_c$ symmetry there is
tunneling between the three deconfined phases which averages the Polyakov loop
to zero. Irrespective of the dynamics, a single quark in a periodic box always
has infinite free energy just for topological reasons. An appropriate
alternative are $C$-periodic boundary conditions which are periodic up to
charge conjugation \cite{Pol91,Kro91}. In that case, the $C$-periodic copy of a
quark is an anti-quark outside the box that can absorb the flux. The 
$C$-periodic box itself contains just a single quark (or more precisely a 
superposition of a quark and an anti-quark) and is thus charged. $C$-periodic 
boundary conditions lead to an explicit breaking of the center symmetry that 
disappears only in the infinite volume limit \cite{Wie92}. As a consequence, 
the Polyakov loop no longer averages to zero and the free energy of a static 
quark becomes finite. In the confined phase the free energy goes to infinity in
the infinite volume limit, while it remains finite in the deconfined phase. 
$C$-periodic boundary conditions are useful in a Gedanken experiment that 
demonstrates the physically observable consequences of the presence of 
deconfined-deconfined domain walls. In the deconfined phase a single quark in a
cubic $C$-periodic volume has a finite free energy. In a long cylindrical 
volume, on the other hand, the free energy increases with the length of the 
cylinder \cite{Hol97}. Hence, the quark is ``confined'' even in the deconfined 
phase, despite the fact that there is no flux string emanating from the quark. 
``Confinement'' in the deconfined phase is a consequence of distinct deconfined
phases being aligned along the long direction of the cylinder. This is an 
indication that spontaneous center symmetry breaking is not just a Euclidean 
artifact. The sensitivity of the quark's free energy to the shape of the 
spatial volume is observable in Minkowski space-time and is characteristic for 
a spontaneously broken $\Z(3)_c$ symmetry. Other interesting subtleties of the 
center symmetry are related to the way it arises in gauge-fixed formulations of
the theory \cite{Len98} and to the 't Hooft loop order parameter \cite{Kor99}.

\subsection{Organization of the Rest of the Article}
 
The rest of this article contains a detailed and more mathematical discussion 
of the topics outlined in the introduction. Section 2 describes the bulk 
physics of gluons at high temperature. In particular, effective theories for 
the Polyakov loop are constructed at non-zero temperature and at non-zero 
chemical potential for static quarks. The nature of confined and deconfined 
phases is discussed and the transitions between them are characterized. In 
section 3 we study the properties and the dynamics of interfaces separating 
coexisting phases of confined and deconfined gluons and gluinos. In particular,
the universal phenomenon of complete wetting is explored using effective 
theories for the Polyakov loop. Section 4 deals with static quarks at finite 
volume and the paradoxical ``confinement'' in the deconfined phase. Finally, 
section 5 contains our conclusions. 

\section{Gluons at High Temperatures}

In this section we introduce $SU(N)$ pure Yang-Mills gauge theory at non-zero
temperature. We discuss the problem in the language of continuum field theory 
but ignore the subtleties that arise in the regularization of ultraviolet 
divergences. Consequently, the resulting expressions are sometimes rather 
formal. We are interested in non-perturbative questions that can in general not
be fully addressed directly in the continuum. Using the lattice regularization,
it is straightforward to replace the formal quantities by completely well 
defined mathematical expressions. We prefer not to do this here because 
mathematical rigor might obscure the basic physics that can be well understood 
in continuum field theory language. Still, we like to emphasize that our
understanding of the center symmetry and its spontaneous breakdown is to a
large extent due to studies in lattice field theory --- the most powerful tool
for understanding non-perturbative field theory from first principles.

\subsection{Path Integral and Euclidean Action}

Let us consider a system of gluons at a non-zero temperature $T = 1/\beta$. Its
partition function is given by
\begin{equation}
\label{Z}
Z = \mbox{Tr} \ [\exp(- \beta H) P].
\end{equation}
Here $H$ is the Yang-Mills Hamilton operator and $P$ is a projection operator
that enforces the Gauss law. It reduces the trace over all states to those that
are gauge invariant and hence physical. We note that $\exp(- \beta H)$ is 
similar to the time-evolution operator $\exp(- i H t)$, except that $\beta$ 
plays the role of an imaginary (or Euclidean) time interval $it$. However, in
order to motivate path integrals in Euclidean time, there is no need to start 
out in real Minkowski time and then Wick-rotate. Our starting point is 
equilibrium thermodynamics which makes no reference to real time evolution and 
puts us directly into Euclidean space-time. One can write a path integral 
expression for the partition function,
\begin{equation}
Z = \int DA \ \exp(- S[A]),
\end{equation}
where 
\begin{equation}
\label{action}
S[A] = \frac{1}{2 g^2} \int_0^\beta dx_4 \int d^3x \ 
\mbox{Tr} F_{\mu\nu} F_{\mu\nu}
\end{equation}
is the Euclidean action. Here $g$ is the gauge coupling and 
\begin{equation}
F_{\mu\nu} = \partial_\mu A_\nu - \partial_\nu A_\mu + [A_\mu,A_\nu]
\end{equation}
is the field strength resulting from the anti-Hermitian non-Abelian vector 
potential
\begin{equation}
A_\mu = i A_\mu^a \lambda^a.
\end{equation}
The $SU(N)$ gluon field is described by $N^2-1$ real-valued functions 
$A_\mu^a(x)$ (with $a \in \{1,2,...,N^2-1\}$) of the Euclidean space-time 
points $x = (\vec x,x_4)$. The $\lambda^a$ are the $N^2-1$ Hermitian generators
of the $SU(N)$ algebra in the fundamental representation. As a consequence of 
the trace in eq.(\ref{Z}), the measure $DA$ of the path integral contains 
fields that obey periodic boundary conditions,
\begin{equation}
\label{boundary}
A_\mu(\vec x,x_4 + \beta) = A_\mu(\vec x,x_4),
\end{equation}
in the Euclidean time direction. In the path integral the projection operator
$P$ that enforces the Gauss law manifests itself as the integration over the
time-like vector potentials $A_4$.

\subsection{Gauge Invariance and Center Symmetry}

The action of eq.(\ref{action}) is invariant under gauge transformations
\begin{equation}
^g\!A_\mu = g (A_\mu + \partial_\mu) g^\dagger,
\end{equation}
with $g(x) \in SU(N)$, under which the field strength transforms as
\begin{equation}
^g\!F_{\mu\nu} = g F_{\mu\nu} g^\dagger.
\end{equation}
In order to maintain the boundary condition eq.(\ref{boundary}) for the vector
potential, we first consider gauge transformations that are strictly periodic 
in Euclidean time
\begin{equation}
g(\vec x,x_4 + \beta) = g(\vec x,x_4).
\end{equation}
Every physical quantity must be invariant under these gauge transformations.
However, in addition there are topologically non-trivial transformations 
\cite{tHo78} that are periodic up to a constant twist matrix $h \in SU(N)$
\begin{equation}
\label{twist}
g(\vec x,x_4 + \beta) = h g(\vec x,x_4).
\end{equation}
When such a transformation is applied to a strictly periodic vector potential 
$A_\mu$ it turns into
\begin{eqnarray}
^g\!A_\mu(\vec x,x_4 + \beta)&=&g(\vec x,x_4 + \beta)
(A_\mu(\vec x,x_4 + \beta) + \partial_\mu) g(\vec x,x_4 + \beta)^\dagger 
\nonumber \\
&=&h g(\vec x,x_4) (A_\mu(\vec x,x_4) + \partial_\mu) 
g(\vec x,x_4)^\dagger h^\dagger \nonumber \\
&=&h \ ^g\!A_\mu(\vec x,x_4) h^\dagger.
\end{eqnarray}
The gauge transformed vector potential $^g\!A_\mu$ obeys the boundary condition
eq.(\ref{boundary}) only if
\begin{equation}
h \ ^g\!A_\mu(\vec x,x_4) h^\dagger = \ ^g\!A_\mu(\vec x,x_4),
\end{equation}
which is the case only if $h$ commutes with $^g\!A_\mu$. This limits us to 
twist matrices $h$ in the center $\Z(N)_c$ of the gauge group $SU(N)$. The 
elements of the center commute with all group elements and are multiples of the
unit matrix
\begin{equation}
h = z \1, \ z = \exp(2 \pi i n/N), \ n \in \{1,2,...,N\}.
\end{equation}
It is important to note that the $\Z(N)$ symmetry gets explicitly broken in the
presence of dynamical fields that transform in the fundamental representation 
of $SU(N)$. For example, quark fields transform as
\begin{equation}
^g\!\Psi = g \Psi,
\end{equation}
and are anti-periodic in Euclidean time,
\begin{equation}
\label{fermionboundary}
\Psi(\vec x,x_4 + \beta) = - \Psi(\vec x,x_4).
\end{equation}
Hence, under a twisted transformation they transform into
\begin{equation}
^g\!\Psi(\vec x,x_4 + \beta) = g(\vec x,x_4 + \beta) \Psi(\vec x,x_4 + \beta) =
- z g(\vec x,x_4) \Psi(\vec x,x_4) = - z \ ^g\!\Psi(\vec x,x_4).
\end{equation}
To maintain the boundary condition eq.(\ref{fermionboundary}) one must restrict
oneself to $z = 1$ so that the center symmetry disappears.

The twisted transformations of eq.(\ref{twist}) represent a global $\Z(N)_c$ 
symmetry of the action of eq.(\ref{action}). Although the action is $\Z(N)_c$ 
invariant, other physical quantities need not necessarily be invariant under 
the center symmetry transformations. In particular, unlike the local $SU(N)$ 
gauge symmetry, the global $\Z(N)_c$ symmetry can break spontaneously and 
should not be considered as a subgroup of the local gauge group. It should be 
pointed out that the center symmetry is a symmetry of the Euclidean action, not
of the Hamiltonian. While the Hamiltonian acts at a given instant in time, the 
$\Z(N)_c$ transformations are characterized by non-trivial boundary conditions 
in Euclidean time and thus affect the entire time evolution. Hence, it is 
meaningless to ask how states in the physical Hilbert space transform under the
center symmetry. Instead, the center symmetry is a symmetry of the spatial 
transfer matrix --- the analog of the Hamiltonian that describes the evolution 
of the system in a spatial direction. The spatial transfer matrix also acts on 
a Hilbert space (however, not on the usual one) and the states in that space 
have definite transformation properties under $\Z(N)_c$. In fact, at non-zero 
temperature the spatial transfer matrix is at least as useful as the 
Hamiltonian. While the spectrum of the Hamiltonian is by definition independent
of temperature, the spectrum of the transfer matrix is sensitive to the
temperature and, for example, contains valuable information about phase 
transitions.

\subsection{The Polyakov Loop as an Order Parameter for Deconfinement}

As we have seen, dynamical fields that transform in the fundamental 
representation break the $\Z(N)_c$ symmetry explicitly. Here we limit ourselves
to cases where such fields are not present. Still, even then one can use static
fundamental charges to probe the physics of the adjoint dynamical fields. 
Static fundamental charges (infinitely heavy test quarks) are described by the
Polyakov loop (a Wilson loop closed around the periodic Euclidean time 
direction) \cite{Pol78,Sus79}
\begin{equation}
\label{Polyakov}
\Phi(\vec x) = \mbox{Tr} \ {\cal P} \exp(\int_0^\beta dx_4 \ A_4(\vec x,x_4)).
\end{equation}
Here ${\cal P}$ denotes path ordering of the exponential. The Polyakov loop is
a complex scalar field that depends on the spatial position $\vec x$ of the 
static color source. In the special case of two colors ($N = 2$) the Polyakov 
loop is always real. The Polyakov loop transforms non-trivially under twisted
$\Z(N)_c$ transformations
\begin{eqnarray}
^g\!\Phi(\vec x)&=&\mbox{Tr} \ {\cal P} 
\exp(\int_0^\beta dx_4 \ ^g\!A_4(\vec x,x_4)) \nonumber \\
&=&\mbox{Tr} \ [g(\vec x,x_4 + \beta) {\cal P} 
\exp(\int_0^\beta dx_4 \ A_4(\vec x,x_4)) g(\vec x,x_4)^\dagger] \nonumber \\
&=&\mbox{Tr} \ [z g(\vec x,x_4) {\cal P}
\exp(\int_0^\beta dx_4 \ A_4(\vec x,x_4)) g(\vec x,x_4)^\dagger] = 
z \Phi(\vec x).
\end{eqnarray}
On the other hand, it is invariant under strictly periodic gauge 
transformations (with $z = 1$) as it should as a gauge invariant physical 
quantity.

The partition function of a system of gluons in the presence of a static
infinitely heavy test quark is given by
\begin{equation}
\label{Zquark}
Z_Q = \int DA \ \Phi(\vec x) \exp(- S[A]).
\end{equation}
The thermal expectation value of the Polyakov loop, 
\begin{equation}
\langle \Phi \rangle = \frac{1}{Z} \int DA \ \Phi(\vec x) \exp(- S[A]) = 
\frac{Z_Q}{Z} = \exp(- \beta F),
\end{equation}
is the ratio of the partition functions of the gluon systems with and without 
the external color source and hence measures the free energy $F$ of the 
external static quark. At low temperatures color is confined and the free
energy of a single quark is infinite ($F = \infty$). Consequently, in the
confined phase $\langle \Phi \rangle = 0$. At high temperatures, on the other
hand, asymptotic freedom suggests that quarks and gluons become deconfined.
Then $F$ is finite and $\langle \Phi \rangle = \Phi_0 \neq 0$ in the deconfined
phase. Since $\Phi$ transforms non-trivially under center symmetry 
transformations, a non-zero expectation value $\Phi_0$ implies that the 
$\Z(N)_c$ symmetry is spontaneously broken in the high-temperature deconfined
phase. It is unusual that a symmetry that is intact at low temperatures gets
spontaneously broken at high temperatures. However, we should keep in mind that
the center symmetry is a symmetry of the spatial transfer matrix --- not a 
symmetry of the Hamiltonian. In fact it is easy to understand why the center
symmetry must break spontaneously in the high-temperature limit $\beta
\rightarrow 0$. In this limit the integral in eq.(\ref{Polyakov}) extends over
shorter and shorter Euclidean time intervals and hence $\Phi_0 \rightarrow 
\mbox{Tr} \1 = N$.

\subsection{Effective Actions for the Polyakov Loop}

To investigate the deconfinement phase transition it is instructive to
consider effective theories for the Polyakov loop. In principle, one can 
imagine to integrate out the gluon fields and derive the effective action for 
the Polyakov loop directly from the underlying Yang-Mills theory
\begin{equation}
\exp(- S[\Phi]) = \int DA \
\delta[\Phi(\vec x) - 
\mbox{Tr} \ {\cal P} \exp(\int_0^\beta dx_4 \ A_4(\vec x,x_4))] \exp(- S[A]).
\end{equation}
Here the $\delta$-functional ensures that $\Phi$ obeys eq.(\ref{Polyakov}). The
resulting effective action is nonlocal and impossible to compute in practice.
Fortunately this is not even necessary. To gain some qualitative insight into 
the nature of the deconfinement phase transition it is sufficient to construct 
simple effective theories that have the desired symmetry properties and to
study their properties. This approach is particularly fruitful when one wants 
to study a universal phenomenon that is common to the Yang-Mills theory and to 
much simpler models. All universal aspects of this phenomenon can then be 
studied in a simple effective theory for the Polyakov loop. Universal behavior 
is naturally associated with second order phase transitions. Then a physical 
correlation length becomes infinitely large and the short-distance details of 
the dynamics become irrelevant. What matters are only the dimensionality of 
space and the symmetry properties of the order parameter.

Let us first construct a simple effective theory for the Polyakov loop in 
$SU(2)$ pure gauge theory. Because the trace of an $SU(2)$ matrix is always
real, the Polyakov loop $\Phi(\vec x)$ is then a real-valued field in three
dimensions. It transforms non-trivially under the $\Z(2)_c$ center 
transformations as $^g\!\Phi = z \Phi$ with $z = \pm 1$. Consequently, it is 
natural to consider the action
\begin{equation}
S[\Phi] = \int d^3x \ [\frac{1}{2} \partial_i \Phi \partial_i \Phi + V(\Phi)]
\end{equation}
with a $\Z(2)_c$ invariant potential
\begin{equation}
V(\Phi) = a \Phi^2 + c \Phi^4.
\end{equation}
Stability of the problem requires $c > 0$. For $a > 0$ the potential has a 
single minimum at $\Phi = 0$ that corresponds to the low-temperature confined
phase. For $a < 0$, on the other hand, there are two degenerate minima at
$\Phi = \pm \Phi_0 = \pm \sqrt{a/2c}$ corresponding to two high-temperature
deconfined phases. The phase transition happens at $a = 0$ and then $\Phi_0
\rightarrow 0$ indicating a second order phase transition. Hence, for $SU(2)$ 
pure gauge theory one indeed expects universal behavior at the deconfinement 
phase transition \cite{McL81a,Sve82}. A simple lattice model with the same 
spatial dimension and the same $\Z(2)$ symmetry is the 3-dimensional Ising 
model. Indeed, detailed lattice gauge theory studies have shown that the 
deconfinement phase transition of the $SU(2)$ Yang-Mills theory is second order
\cite{McL81b} and in the universality class of the 3-d Ising model 
\cite{Eng90}. Since the order parameter vanishes continuously, at a second 
order phase transition the various phases become indistinguishable at the 
transition. The lattice calculations also give the ratio of the critical
temperature $T_c$ and the square root of the zero-temperature string tension 
$\sigma$ as $T_c/\sqrt{\sigma} = 0.69(1)$ \cite{Fin93}.

For three colors ($N = 3$) the situation is different. First of all, the
Polyakov loop $\Phi = \Phi_1 + i \Phi_2$ is now a complex scalar field with 
real part $\Phi_1$ and imaginary part $\Phi_2$. Consequently, besides the 
$\Z(3)$ center symmetry, we now must also consider the charge conjugation 
symmetry.\footnote{For two colors charge conjugation is equivalent to a global 
gauge transformation.} Charge conjugation replaces the non-Abelian vector 
potential by its complex conjugate
\begin{equation}
^C\!A_\mu = A_\mu^*.
\end{equation}
Accordingly the Polyakov loop also turns into its complex conjugate
\begin{eqnarray}
^C\!\Phi(\vec x)&=&\mbox{Tr} \ 
{\cal P} \exp(\int_0^\beta dx_4 \ ^C A_4(\vec x,x_4)) \nonumber \\
&=&\mbox{Tr} \ {\cal P} \exp(\int_0^\beta dx_4 \ A_4(\vec x,x_4)^*) = 
\Phi(\vec x)^*.
\end{eqnarray}
The natural candidate for an effective action now is
\begin{equation}
\label{effectiveaction}
S[\Phi] = \int d^3x \ [\frac{1}{2} \partial_i \Phi^* \partial_i \Phi + 
V(\Phi)],
\end{equation}
with a $\Z(3)_c$ and charge conjugation invariant potential, i.e.
\begin{equation}
V(z \Phi) = V(\Phi), \ V(\Phi^*) = V(\Phi).
\end{equation}
The most general quartic potential that obeys these symmetries is given by
\begin{equation}
\label{potential}
V(\Phi) = a |\Phi|^2 + b \Phi_1(3 \Phi_2^2 - \Phi_1^2) + c |\Phi|^4.
\end{equation}
Again, in order to have a stable system we need $c > 0$. For $a > 0$ the system
has a minimum at $\Phi_1 = \Phi_2 = 0$ corresponding to the confined phase. For
$0 < a < 9 b^2/32 c$ there is another minimum at
\begin{equation}
\Phi_1 = \Phi_0 = \frac{3 b + \sqrt{9 b^2 - 32 a c}}{8 c}, \ \Phi_2 = 0,
\end{equation}
which has two other $\Z(3)_c$ copies. We denote the four minima by
\begin{equation}
\Phi^{(1)} = \Phi_0, \ \Phi^{(2)} = \frac{1}{2}(- 1 + i \sqrt{3}) \Phi_0, \
\Phi^{(3)} = \frac{1}{2}(- 1 - i \sqrt{3}) \Phi_0, \ \Phi^{(4)} = 0.
\end{equation}
The first three minima correspond to the three high-temperature deconfined
phases. For $b^2 = 4 a c$ all four minima are degenerate. This corresponds to
the phase transition temperature. However, in contrast to the $SU(2)$ case, at
the phase transition we now have $\Phi_0 = \sqrt{a/c} = b/2c \neq 0$ which 
corresponds to a first order phase transition. At a first order phase 
transition one does not have universal behavior in the bulk because the bulk 
correlation length remains finite and the short distance details of the 
dynamics remain important at the phase transition. A simple lattice model in 
the same spatial dimension and with the same global symmetries as the $SU(3)$ 
pure gauge theory is the 3-d 3-state Potts model \cite{Wu82} which indeed has 
a first order phase transition \cite{Kna79}. Furthermore, detailed numerical 
studies in lattice gauge theory have shown that the deconfinement phase 
transition in $SU(3)$ Yang-Mills theory is also first order \cite{Cel83}. In
this case the ratio of the critical temperature and the square root of the
string tension is $T_c/\sqrt{\sigma} = 0.64(1)$.

\subsection{Gluons in a Background of Static Quarks}

As we have seen, the partition function of a system of gluons in the presence
of a single static quark at position $\vec x$ is given by eq.(\ref{Zquark}).
Similarly, the gluon partition function in the presence of a single anti-quark 
is
\begin{equation}
Z_{\overline Q} = \int DA \ \Phi(\vec x)^* \exp(- S[A]).
\end{equation}
Consequently, the partition function of a system of gluons in the background of
$n$ static quarks and $\overline n$ static anti-quark takes the form
\begin{equation}
Z_{n,\overline n} = \int DA \ \frac{1}{n!} [\int d^3x \ \Phi(\vec x)]^n 
\frac{1}{\overline n!} [\int d^3y \ \Phi(\vec y)^*]^{\overline n} \exp(- S[A]).
\end{equation}
Here we have integrated over the positions $\vec x$ of quarks and $\vec y$ of
anti-quarks. The factors $n!$ and $\overline n!$ take into account that quarks
as well as anti-quarks are indistinguishable particles. We consider quarks of 
mass $M$ with quark chemical potential $\mu$ that couples to the baryon number 
$B = (n - \overline n)/3$ and thus obtain
\begin{eqnarray}
Z(\mu)&=&\sum_{n,\overline n} Z_{n,\overline n} 
\exp(- \beta M(n + \overline n)) \exp(\beta \mu(n - \overline n)) \nonumber \\
&=&\int DA \sum_{n,\overline n} \frac{1}{n!} [\int d^3x \ \Phi(\vec x)]^n
\exp(- \beta (M - \mu) n) \nonumber \\ 
&\times&\frac{1}{\overline n!} [\int d^3y \ \Phi(\vec y)^*]^{\overline n} 
\exp(- \beta (M + \mu) \overline n) \exp(- S[A]) \nonumber \\
&=&\int DA \exp[\int d^3x \ \Phi(\vec x) \exp(- \beta (M - \mu))] \nonumber \\
&\times&\exp[\int d^3y \ \Phi(\vec y)^* \exp(- \beta (M + \mu))] \exp(- S[A]).
\end{eqnarray}
Consequently, the effective action for the gluons now takes the form
\begin{equation}
S_{eff}[A] = S[A] - \int d^3x \ [\Phi(\vec x) \exp(- \beta (M - \mu)) +
\Phi(\vec x)^* \exp(- \beta (M + \mu))].
\end{equation}
As expected in the presence of fundamental charges, the additional terms in the
effective action break the center symmetry explicitly. In the presence of the
chemical potential, the effective action is in general complex. The $SU(2)$ 
case is exceptional because then the Polyakov loop and hence the effective
action take real values. The action is real even in the $SU(N)$ case for purely
imaginary values of the chemical potential. In that case, $\mu = i B_4$ enters 
the action as a constant vector potential $B_4$ in the 4-direction that couples
to the baryon number. A real chemical potential, on the other hand, enters the 
action as a purely imaginary vector potential. Taking the limit $M, \mu 
\rightarrow \infty$ with $M - \mu$ finite we obtain
\begin{equation}
S_{eff}[A] = S[A] - \int d^3x \ \Phi(\vec x) \exp(- \beta (M - \mu)).
\end{equation}
The chemical potential favors positive real values of $\Phi$. 

In the simple $\Phi^4$ model for the $SU(3)$ Polyakov loop the chemical 
potential enters the effective potential as
\begin{equation}
\label{pot1}
V(\Phi) = a |\Phi|^2 + b \Phi_1(3 \Phi_2^2 - \Phi_1^2) + c |\Phi|^4 - h \Phi,
\end{equation}
where $h$ represents $\exp(- \beta (M - \mu))$. At non-zero $h$ the deconfined
phase $d^{(1)}$ with a real valued Polyakov loop is favored while the other two
deconfined phases are suppressed. The deconfined phase $d^{(1)}$ coexists with
the confined phase along a first order transition line that terminates in a
critical endpoint. Since the $\Z(3)_c$ symmetry is explicitly broken in the
presence of the chemical potential, at non-zero $h$ the Polyakov loop is
non-zero even in the confined phase. We denote the values of the Polyakov loop
in the confined and the deconfined phase by $\Phi^{(c)}$ and $\Phi^{(d)}$,
respectively. Along the first order transition line at which the confined and
the deconfined phase coexist, the potential (at $\Phi_2 = 0$) can then be 
written as
\begin{equation}
\label{pot2}
V(\Phi) = c(\Phi_1 - \Phi^{(d)})^2 (\Phi_1 - \Phi^{(c)})^2 + V_0.
\end{equation}
Comparing eq.(\ref{pot1}) and eq.(\ref{pot2}) one derives a condition for the
location of the first order phase transition line
\begin{equation}
\label{line}
\frac{4ac}{b^2} - 1 = \frac{8c^2}{b^3} h.
\end{equation}
It is non-trivial to analyze the phase diagram with lattice field theory
methods because the complex action prevents the use of standard importance
sampling techniques \cite{Blu96}. Using a cluster algorithm the complex action 
problem has been solved for the 3-d 3-state Potts model \cite{Alf00}. The 
corresponding phase diagram is shown in figure 1. 
\begin{figure}
\begin{center}
\epsfig{file=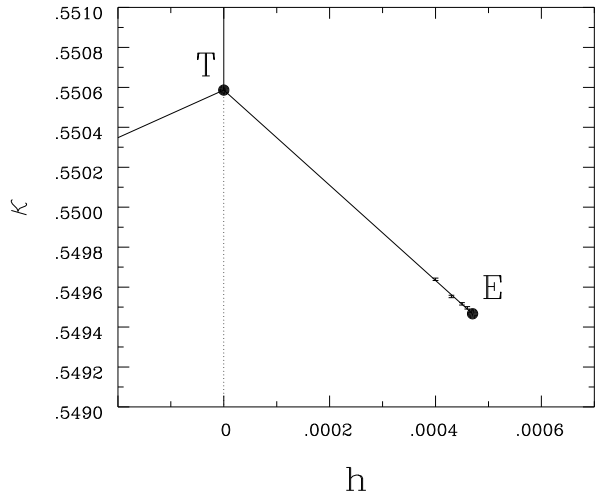,
width=6.5cm,angle=90,
bbllx=50,bblly=200,bburx=535,bbury=800}
\end{center}
\caption{\it The phase diagram of the 3-d 3-state Potts model in the 
$(h,\kappa)$-plane. Qualitatively, $\kappa$ plays the role of the temperature
in Yang-Mills theory. The ordinary deconfinement phase transition at $h = 0$ is
a triple point $T$ from which a line of first order phase transitions emerges. 
This line terminates in the critical endpoint $E$.}
\end{figure}
In the Potts model the parameter $\kappa - \kappa_c$ plays the role of 
$1 - 4ac/b^2$. Interestingly, the transition line in the Potts model is 
consistent with a straight line as predicted by eq.(\ref{line}). The 
discontinuity of the Polyakov loop across the transition takes the form
\begin{equation}
\label{gap}
\Phi^{(d)} - \Phi^{(c)} = \frac{b}{2c} \sqrt{1 - \frac{h}{h_E}}, \ 
h_E = \frac{b^3}{16c^2}.
\end{equation}
It vanishes at the critical endpoint $h = h_E$ in which the first order line
terminates at a second order phase transition. A critical exponent $\beta$
(not to be confused with the inverse temperature) is defined by
\begin{equation}
\Phi^{(d)} - \Phi^{(c)} \propto (1 - \frac{h}{h_E})^\beta.
\end{equation}
The simple mean field theory result of eq.(\ref{gap}) yields $\beta = 1/2$. 
However, this result is not reliable. Instead, one expects the endpoint to be 
in the universality class of the 3-d Ising model which has $\beta = 0.3260(8)$.
A similar endpoint has been studied numerically by Karsch and Stickan in the
3-d 3-state Potts model with a real action \cite{Kar00}. Indeed they find very
good agreement with universal 3-d Ising behavior.

\section{Interfaces Separating Confined and Deconfined Phases}

At a first order phase transition the various phases remain distinguishable 
because the order parameter is discontinuous at the transition 
($\Phi_0 \neq 0$). Consequently, at a first order phase transition distinct 
phases coexist with one another. The phases are spatially separated from each 
other by interfaces with a non-zero free energy per unit area (the interface 
tension). In $SU(3)$ Yang-Mills theory there are confined-deconfined interfaces
that separate the low and high-temperature phases which coexist at the first 
order deconfinement phase transition. In $SU(2)$ gauge theory, on the other 
hand, such interfaces do not exist because the confined and the deconfined 
phase become indistinguishable at the second order phase transition. Both in 
$SU(2)$ and $SU(3)$ Yang-Mills theory there are deconfined-deconfined domain 
walls that separate the various deconfined phases at high temperatures. When a 
deconfined-deconfined domain wall in the $SU(3)$ theory is cooled down to the 
phase transition, it splits into a pair of confined-deconfined interfaces. Then
the confined phase forms a complete wetting layer that grows to a macroscopic 
size at the phase transition. At the phase transition the size of the wetting 
layer diverges and again the microscopic details of the dynamics become 
irrelevant.

\subsection{Deconfined-Deconfined Domain Walls and Confined-Deconfined
Interfaces}

When a discrete symmetry gets spontaneously broken, there exist a number of 
degenerate phases related to one another by symmetry transformations. These
phases can coexist and are separated by domain walls. In the gluon plasma the
spontaneous breakdown of the $\Z(3)_c$ center symmetry gives rise to three
distinct deconfined phases. These phases are separated from one another by
deconfined-deconfined domain walls. The domain walls are stable at all
temperatures above $T_c$. Due to asymptotic freedom, the domain wall tension 
(free energy per unit area) can be computed semi-classically in the 
high-temperature limit. The domain wall is a soliton in the effective potential
for the Polyakov loop. The 1-loop effective potential for the complex phase of 
the Polyakov loop was first calculated by Weiss \cite{Wei81}. This calculation
shows that there are three degenerate minima corresponding to three distinct
deconfined phases. Based on this result, Bhattacharya, Gocksch, Korthals Altes 
and Pisarski have analytically determined the tension of a 
deconfined-deconfined domain wall in the high-temperature limit \cite{Bha91}
\begin{equation}
\alpha_{dd} = \frac{8 \pi^2}{9 g} T^3,
\end{equation}
where $g$ is the gauge coupling renormalized at the scale $T$.

Interfaces separating coexisting phases also occur at any first order phase 
transition. In that case, the coexisting phases are in general not related by 
symmetry. For example, in $SU(3)$ Yang-Mills theory the high-temperature gluon 
plasma coexists with the low-temperature glueball phase at $T_c$. The 
corresponding confined-deconfined interfaces are stable only at $T_c$. Unlike 
the deconfined-deconfined domain wall tension, the confined-deconfined 
interface tension cannot be computed semi-classically since $T_c$ is too small 
for perturbation theory to be applicable. Instead there have been several
attempts to extract the interface tension from lattice calculations 
\cite{Hua90}. It should be noted that these calculations do not contain 
dynamical quark effects. Due to rather strong lattice spacing artefacts, it is 
difficult to reliably extrapolate the results to the continuum limit. A rough 
estimate of the confined-deconfined interface tension gives $\alpha_{cd} 
\approx 0.020(5) T_c^3$. This result is interesting from a cosmological point 
of view. The confined-deconfined interface tension sets a scale for the 
nucleation rate of bubbles of confined phase from the quark-gluon plasma and 
hence for the spatial inhomogeneities generated at temperatures around 100 MeV.
These inhomogeneities may have an effect on the primordial synthesis of light 
atomic nuclei which takes place when the universe has cooled down to 
temperatures in the 1 MeV range. The small value for $\alpha_{cd}$ from above
suggests that the inhomogeneities are small and that the standard scenario of
primordial nucleosynthesis is correct. 

\subsection{Complete Wetting of Hot Gluons}

Frei and Patk\'{o}s were first to conjecture that complete wetting occurs in 
$SU(3)$ pure gauge theory at the high-temperature deconfinement phase 
transition \cite{Fre89}. The interface tensions $\alpha_{cd}$ and $\alpha_{dd}$
of confined-deconfined interfaces and deconfined-deconfined domain walls, 
respectively, determine the shape of a droplet of confined phase that wets a 
deconfined-deconfined domain wall. As shown in figure 2a, such a droplet forms 
a lens with opening angle $\theta$, where
\begin{equation}
\alpha_{dd} = 2 \alpha_{cd} \cos{\frac{\theta}{2}}.
\end{equation}
\begin{figure}[t]
\psfig{figure=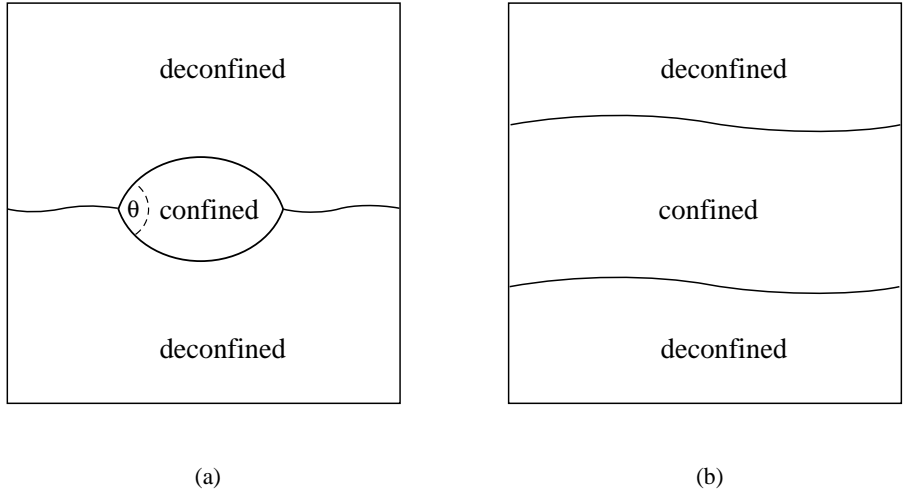,height=2.56in,width=4.7in}
\caption{\it Incomplete versus complete wetting. (a) For $\alpha_{dd} < 
2 \alpha_{cd}$ one has incomplete wetting with $\theta \neq 0$. Then the 
confined phase forms a lens shaped droplet at the deconfined-deconfined domain
wall. (b) Complete wetting corresponds to $\alpha_{dd} = 2 \alpha_{cd}$. Then
$\theta = 0$, and the confined phase forms a film that splits the
deconfined-deconfined domain wall into two confined-deconfined interfaces.} 
\end{figure}
This equation for $\theta$ follows from the forces at the corner of the lens 
being in equilibrium. In condensed matter physics 
the inequality 
\begin{equation}
\alpha_{dd} \leq 2 \alpha_{cd},
\end{equation}
was derived by Widom \cite{Wid75}. It follows from thermodynamic stability 
because a hypothetical deconfined-deconfined domain wall with  
$\alpha_{dd} > 2 \alpha_{cd}$ would simply split into two confined-deconfined 
interfaces. If $\alpha_{dd} < 2 \alpha_{cd}$ the confined phase forms droplets 
at a deconfined-deconfined domain wall. In condensed matter physics this 
phenomenon is known as incomplete wetting. For $\alpha_{dd} = 2 \alpha_{cd}$, 
on the other hand, $\theta = 0$ and the lens shaped droplet degenerates to an 
infinite film, as shown in figure 2b. When such a film is formed, this is 
called complete wetting.

Complete wetting is a universal phenomenon of interfaces characterized by
several critical exponents. For example, the width $r$ of the complete
wetting layer diverges as
\begin{equation}
r \propto (T - T_c)^{-\psi},
\end{equation}
where $T - T_c$ measures the deviation from the point of phase coexistence and
$\psi$ is a critical exponent. The value of $\psi$ depends on the range of the
interactions between the two interfaces that enclose the complete wetting
layer. In $SU(3)$ Yang-Mills theory the interactions between the interfaces are
short-ranged and the interaction energy per unit area is given by 
$\gamma \exp(- r/r_0)$. In addition, above $T_c$ the confined phase that 
forms the wetting layer has a slightly larger bulk free energy than the 
deconfined phases. Close to $T_c$, the additional bulk free energy per unit 
area for a wetting layer of width $r$ is given by $\delta (T - T_c) r$. Hence, 
the total free energy per unit area of the two interface system relative to the
free energy of two infinitely separated interfaces at $T_c$ is given by
\begin{equation}
\alpha_{dd}(T) - 2 \alpha_{cd}(T_c) = \gamma \exp(- r/r_0) + 
\delta (T - T_c) r.
\end{equation}
Minimizing the free energy with respect to $r$, one finds the equilibrium width
\begin{equation}
r = - r_0 \log \frac{\delta r_0}{\gamma}(T - T_c),
\end{equation}
such that $\psi = 0$. Inserting the equilibrium value of $r$ one finds
\begin{equation}
\label{tensions}
\alpha_{dd}(T) - 2 \alpha_{cd}(T_c) = \delta r_0 (T - T_c) 
[1 - \log \frac{\delta r_0}{\gamma}(T - T_c)].
\end{equation}
This is consistent with the critical behavior in the $\Z(3)_c$ symmetric 
$\Phi^4$ theory \cite{Tra92}. As we will see later, the same critical exponents
follow for the high-temperature deconfinement transition in the supersymmetric 
case.

The universal aspects of the dynamics of confined-deconfined interfaces and
deconfined-deconfined domain walls in $SU(3)$ Yang-Mills theory can be 
investigated using the effective action for the Polyakov loop of 
eq.(\ref{effectiveaction}) \cite{Tra92}. The corresponding potential takes the 
form $V(\Phi) = a |\Phi|^2 + b \Phi_1 (3 \Phi_2^2 - \Phi_1^2) + c |\Phi|^4$ of
eq.(\ref{potential}). The deconfinement phase transition occurs at $b^2 = 4ac$
because then the confined phase minimum at $\Phi^{(4)} = 0$ is degenerate with 
the deconfined phase minimum at $\Phi^{(1)} = \Phi_0$ and its $\Z(3)_c$ copies 
at $\Phi^{(2)}$ and $\Phi^{(3)}$. The profile $\Phi(\vec x) = \Phi(z)$ of a 
planar interface perpendicular to the $z$-direction follows from the solution 
of the corresponding classical equations of motion
\begin{eqnarray}
\label{eqofmotion}
&&\frac{d^2 \Phi_1}{dz^2} = \frac{\partial V}{\partial \Phi_1} = 
2 a \Phi_1 + 3 b (\Phi_2^2 - \Phi_1^2) + 4 c \Phi_1 (\Phi_1^2 + \Phi_2^2),
\nonumber \\
&&\frac{d^2 \Phi_2}{dz^2} = \frac{\partial V}{\partial \Phi_2} = 
2 a \Phi_2 + 6 b \Phi_1 \Phi_2 + 4 c \Phi_2 (\Phi_1^2 + \Phi_2^2).
\end{eqnarray}
The solution for a confined-deconfined interface interpolating between the
confined phase at negative $z$ and the deconfined phase $\Phi^{(1)}$ at 
positive $z$ takes the form
\begin{equation}
\Phi_1(z) = \frac{1}{2} \Phi_0 [1 + \tanh(\mu z)], \ \Phi_2(z) = 0.
\end{equation}
Here $\mu = b/\sqrt{8c}$ determines the width of the interface. The 
confined-deconfined interface tension $\alpha_{cd}$ is given by
\begin{equation}
\beta \alpha_{cd} = \int_{- \infty}^\infty dz \ 
[\frac{1}{2} \partial_i \Phi^* \partial_i \Phi + V(\Phi)] =
2 \int_{- \infty}^\infty dz \ V(\Phi) = \frac{\mu \Phi_0^2}{3}.
\end{equation}
The interface profile, its width, as well as its tension are non-universal
properties that are quantitatively different in the simple effective $\Phi^4$
theory and in $SU(3)$ Yang-Mills theory.

Still, there are universal properties related to the phenomenon of complete
wetting. To investigate this, we now consider the profile of a 
deconfined-deconfined domain wall that interpolates between the deconfined
phases $\Phi^{(2)}$ and $\Phi^{(3)}$. The domain wall profile is a solution of
eq.(\ref{eqofmotion}) with the condition that $\Phi$ approaches $\Phi^{(2)}$ at
$z = \infty$ and $\Phi^{(3)}$ at $z = - \infty$. A typical solution deep in the
deconfined phase is illustrated in fig.3. There is a sharp transition from one
deconfined phase to the other. Figure 4 shows the domain wall profile very 
close to the deconfinement phase transition. In this temperature region the 
deconfined-deconfined domain wall splits into a pair of confined-deconfined
interfaces that enclose a complete wetting layer of confined phase.
\begin{figure}[t]
\psfig{figure=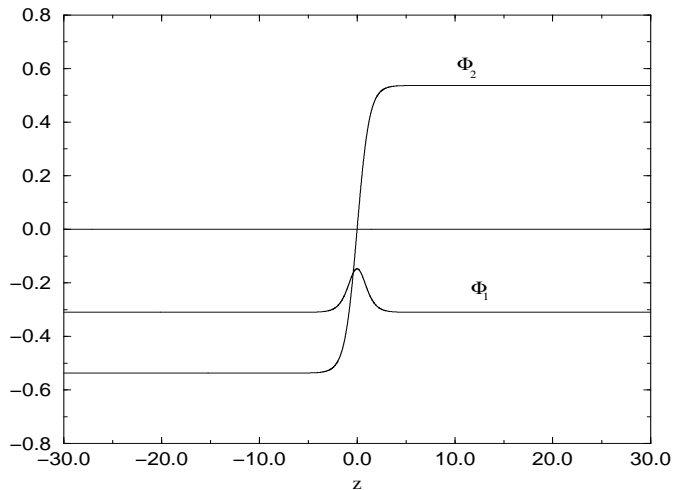,height=3in,width=4.125in}
\caption{\it Profile of a deconfined-deconfined domain wall in $SU(3)$ 
Yang-Mills theory deep in the deconfined phase.}
\end{figure}
\begin{figure}[t]
\psfig{figure=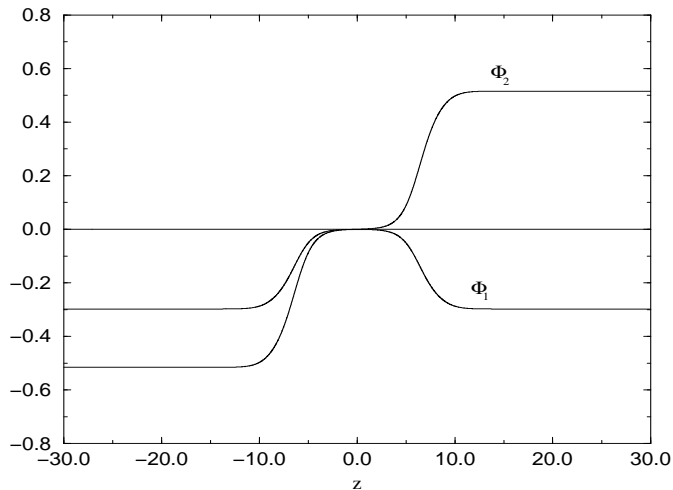,height=3in,width=4.125in}
\caption{\it Profile of a deconfined-deconfined domain wall in $SU(3)$ 
Yang-Mills theory close to the phase transition where the wall splits into two 
confined-deconfined interfaces with a complete wetting layer of confined phase 
between them.}
\end{figure}
Very close to the phase transition one can even construct an analytic solution
of the completely wet deconfined-deconfined domain wall by combining two 
solutions for confined-deconfined interfaces
\begin{eqnarray}
\Phi_1(z)&=&- \frac{1}{4} \Phi_0 [2 + \tanh\mu(z-z_0) - \tanh\mu(z+z_0)],
\nonumber \\
\Phi_2(z)&=&\frac{\sqrt{3}}{4} \Phi_0 [\tanh\mu(z-z_0) + \tanh\mu(z+z_0)].
\end{eqnarray}
The interface separation and hence the width of the confined complete wetting
layer is given by
\begin{equation}
r = 2 z_0 = - \frac{1}{4 \mu} \log(\frac{1}{2} - \frac{2ac}{b^2}).
\end{equation}
Since the phase transition corresponds to $b^2 = 4ac$, the width of the wetting
layer diverges logarithmically at $T_c$. This is consistent with the critical
exponent $\psi = 0$ as expected for interfaces with short-range interactions.
The interface tension of the deconfined-deconfined domain wall is given by
\begin{equation} 
\beta \alpha_{dd} = \mu \Phi_0^2 [\frac{2}{3} - 2(1 - \frac{4ac}{b^2})].
\end{equation}
Hence, consistent with the leading $(T - T_c)$ dependence in 
eq.(\ref{tensions}), one obtains
\begin{equation}
\beta(2 \alpha_{cd} - \alpha_{dd}) = 
2 \mu \Phi_0^2 (1 - \frac{4ac}{b^2}).
\end{equation}

\subsection{Strings Ending on Walls in Supersymmetric Yang-Mills Theory}

Witten has argued that color flux strings in supersymmetric Yang-Mills theory
can end on domain walls \cite{Wit97}. This effect follows from a calculation in
M-theory, where the domain wall is represented by a D-brane on which strings 
can end. In supersymmetric Yang-Mills theory a $\Z(3)_\chi$ chiral symmetry ---
an unbroken remnant of the anomalous $U(1)_R$ symmetry --- is spontaneously 
broken at low temperatures by a non-zero value of the gluino condensate $\chi$.
As a consequence, there are three distinct confined phases, characterized by 
three different values of $\chi$ which are related by $\Z(3)_\chi$ 
transformations. Regions of space filled with different confined phases are 
separated by confined-confined domain walls. The properties of domain walls in 
supersymmetric theories have been investigated in great detail by Shifman and
collaborators \cite{Kov97,Kog98} and also by Carroll, Hellerman and Trodden 
\cite{Car98}. Often, topological defects which result from spontaneous symmetry
breaking have the phase of unbroken symmetry at their cores. For example, 
magnetic monopoles or cosmic strings have symmetric vacuum at their centers. 
Here we construct an effective theory for the Polyakov loop and the gluino 
condensate in which $\Z(3)_\chi$ restoration implies the breaking of the 
$\Z(3)_c$ center symmetry. Then the high-temperature deconfined phase appears 
at the center of the domain wall forming a complete wetting layer. As a 
consequence, the Polyakov loop has a non-zero expectation value there and thus 
a static quark has a finite free energy close to the wall so that its string 
can end there. In supersymmetric Yang-Mills theory strings can end on 
confined-confined domain walls because they can transport color flux to 
infinity at a finite free energy cost. A similar scenario involving a 
non-Abelian Coulomb phase has been discussed at zero temperature \cite{Kog98}.

In supersymmetric $SU(3)$ Yang-Mills theory the $\Z(3)_\chi$ chiral symmetry is
spontaneously broken in the confined phase. The corresponding order parameter 
is the complex valued gluino condensate $\chi = \chi_1 + i \chi_2$. Under 
chiral transformations $z \in \Z(3)_\chi$ the gluino condensate transforms 
into $\chi' = \chi z$ and under charge conjugation it gets replaced by its 
complex conjugate. At high temperatures one expects chiral symmetry to be 
restored and --- as in the non-supersymmetric theory --- the $\Z(3)_c$ center 
symmetry to be spontaneously broken due to deconfinement. Consequently, the 
effective action describing the interface dynamics now depends on both order 
parameters $\Phi$ and $\chi$ and
\begin{equation}
S[\Phi,\chi] = \int d^3 x \ [\frac{1}{2} \partial_i \Phi^* \partial_i \Phi + 
\frac{1}{2} \partial_i \chi^* \partial_i \chi + V(\Phi,\chi)].
\end{equation}
The most general quartic potential consistent with $\Z(3)_c$, $\Z(3)_\chi$ and
charge conjugation symmetry now takes the form
\begin{eqnarray}
V(\Phi,\chi)&=&a |\Phi|^2 + b \Phi_1(3 \Phi_2^2 - \Phi_1^2) + c |\Phi|^4 
\nonumber \\
&+&d |\chi|^2 + e \chi_1(3 \chi_2^2 - \chi_1^2) + f |\chi|^4 + 
g |\Phi|^2 |\chi|^2.
\end{eqnarray}
Assuming that deconfinement and chiral symmetry restoration occur at the same 
temperature and that the phase transition is first order, three chirally broken
confined phases coexist with three distinct chirally symmetric deconfined 
phases. The three deconfined phases have $\Phi^{(1)} = \Phi_0$, 
$\Phi^{(2)} = (-1/2 + i\sqrt{3}/2) \Phi_0$ and $\Phi^{(3)} = 
(-1/2 - i\sqrt{3}/2) \Phi_0$ and $\chi^{(1)} = \chi^{(2)} = \chi^{(3)} = 0$, 
while the three confined phases are characterized by $\Phi^{(4)} = \Phi^{(5)} =
\Phi^{(6)} = 0$ and $\chi^{(4)} = \chi_0 \in \R$, $\chi^{(5)} = 
(-1/2 + i\sqrt{3}/2) \chi_0$ and $\chi^{(6)} = (-1/2 - i\sqrt{3}/2) \chi_0$. 
The phase transition temperature corresponds to a choice of parameters 
$a,b,...,g$ such that all six phases $\Phi^{(n)}$, $\chi^{(n)}$ represent 
degenerate absolute minima of $V(\Phi,\chi)$.

We now look for solutions of the classical equations of motion, representing
planar domain walls, i.e. $\Phi(\vec x) = \Phi(z)$, $\chi(\vec x) = \chi(z)$, 
where $z$ again is the coordinate perpendicular to the wall. The equations of 
motion then take the form
\begin{equation}
\frac{d^2 \Phi_i}{dz^2} = \frac{\partial V}{\partial \Phi_i}, \
\frac{d^2 \chi_i}{dz^2} = \frac{\partial V}{\partial \chi_i}.
\end{equation}
\begin{figure}[t]
\psfig{figure=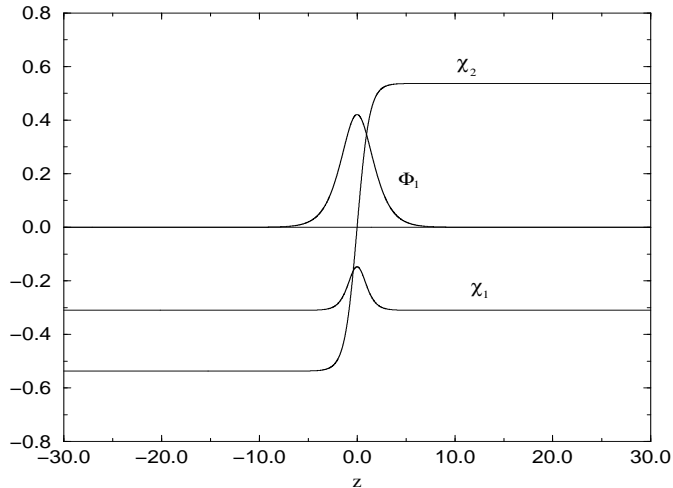,height=3in,width=4.125in}
\caption{\it Profile of a confined-confined domain wall in supersymmetric 
$SU(3)$ Yang-Mills theory deep in the confined phase. Note that 
$\Phi_1(0) \neq 0$, i.e. the center of the wall has properties of the 
deconfined phase.} 
\end{figure}
\begin{figure}[t]
\psfig{figure=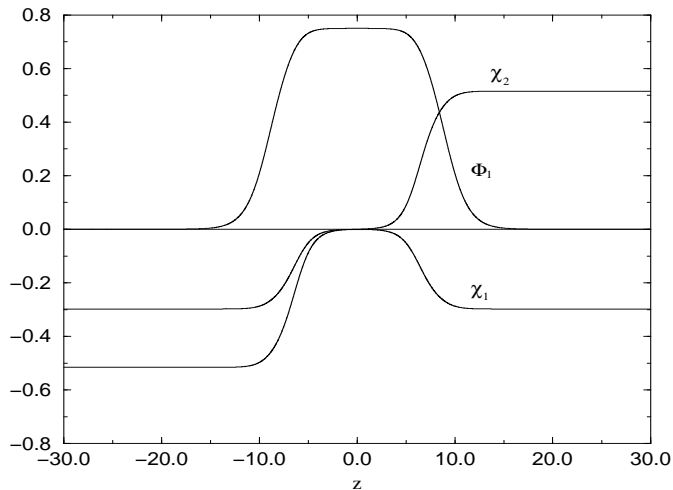,height=3in,width=4.125in}
\caption{\it Profile of a confined-confined domain wall in supersymmetric 
$SU(3)$ Yang-Mills theory close to the phase transition. In this region the 
wall splits into two confined-deconfined interfaces with a complete wetting 
layer of deconfined phase between them.}
\end{figure}
Figure 5 shows a numerical solution of these equations for a domain wall 
separating two confined phases of type $c^{(5)}$ and $c^{(6)}$, i.e. with 
boundary conditions $\Phi(\infty) = \Phi^{(5)}$, $\chi(\infty) = \chi^{(5)}$ 
and $\Phi(-\infty) = \Phi^{(6)}$, $\chi(-\infty) = \chi^{(6)}$. Figure 5 
corresponds to a temperature deep in the confined phase. Still, at the domain
wall the Polyakov loop is non-zero, i.e. the center of the domain wall shows
characteristic features of the deconfined phase. Figure 6 corresponds to
a temperature very close to the phase transition. Then the confined-confined
domain wall splits into two confined-deconfined interfaces and the deconfined
phase forms a complete wetting layer between them. The solutions of figure 5
have deconfined phase of type $d^{(1)}$ at their centers. Due to the $\Z(3)_c$
symmetry, there are related solutions with deconfined phase of types $d^{(2)}$ 
and $d^{(3)}$.

For the special values $d = a = 0$, $e = b$, $f = c$, $g = 2 c$ one can find an
analytic solution for a confined-deconfined interface. Combining two of these
solutions to a confined-confined domain wall, one obtains
\begin{eqnarray}
\Phi_1(z)&=&- \frac{1}{2} \Phi_0 [\tanh\mu(z-z_0) - \tanh\mu(z+z_0)], \ 
\Phi_2(z) = 0, \nonumber \\
\chi_1(z)&=&- \frac{1}{4} \chi_0 [2 + \tanh\mu(z-z_0) - \tanh\mu(z+z_0)],
\nonumber \\
\chi_2(z)&=&\frac{\sqrt{3}}{4} \chi_0 [\tanh\mu(z-z_0) + \tanh\mu(z+z_0)],
\end{eqnarray}
where $\Phi_0 = \chi_0 = 3b/4c$ and $\mu = 3b/4\sqrt{c}$. The critical
temperature corresponds to $e^4/f^3 = b^4/c^3$. Near criticality, where
$\Delta = e^4/f^3 - b^4/c^3$ is small, the above solution is valid up to order
$\Delta^{1/2}$, while now $e = b$ and $f = c$ are satisfied to order $\Delta$.
The width of the deconfined complete wetting layer,
\begin{equation}
r = 2 z_0 = - \frac{1}{2 \mu} \log \Delta + C,
\end{equation}
where $C$ is a constant, grows logarithmically as we approach the phase 
transition temperature. This is the expected critical behavior for interfaces 
with short-range interactions, which again implies the critical exponent
$\psi = 0$.

Now we wish to explain why the appearance of the deconfined phase at the center
of the domain wall allows a QCD string to end there. We recall that an
expectation value $\langle \Phi \rangle \neq 0$ implies that the free energy 
of a static quark is finite. Indeed, the solution of figure 5 containing 
deconfined phase of type $d^{(1)}$ has $\Phi_1(0) \neq 0$ so that a static 
quark located at the center of the wall has finite free energy. As one moves 
away from the wall, the Polyakov loop decreases as
\begin{equation}
\Phi_1(z) \propto \exp(- \beta F(z)) \propto 
\exp(- z \sqrt{2a + 2 g \chi_0^2}).
\end{equation}
Consequently, as the static quark is displaced from the wall, its free energy
$F(z)$ increases linearly with the distance $z$ from the center, i.e. the quark
is confined to the wall. The string emanating from the static quark ends on the
wall and has a tension
\begin{equation}
\sigma = \lim_{z \rightarrow \infty} \frac{F(z)}{z} = 
\frac{1}{\beta} \sqrt{2a + 2 g \chi_0^2}.
\end{equation}
Still, there are the other domain wall solutions (related to the one from above
by $\Z(3)_c$ transformations), which contain deconfined phase of types 
$d^{(2)}$ and $d^{(3)}$. One could argue that, after path integration over all 
domain wall configurations, one obtains $\langle \Phi \rangle = 0$. However, 
this is not true. In fact, the wetting layer at the center of a 
confined-confined domain wall is described by a two-dimensional field theory 
with a spontaneously broken $\Z(3)_c$ symmetry. Consequently, deconfined phase 
of one definite type spontaneously appears at the domain wall. 

\section{Paradoxical ``Confinement'' in the Deconfined Phase} 

We now want to study some subtle effects related to the spontaneous breakdown 
of the center symmetry at high temperatures. Whenever a global symmetry gets
spontaneously broken, the system becomes very sensitive to the spatial
boundary conditions. To expose this sensitivity explicitly, it is often very 
instructive to work in a finite spatial volume. This is much better than 
working in an infinite volume with unspecified boundary conditions, which tends
to obscure the physics in the infrared. Instead, by consistently working in a 
finite volume, we learn how the infinite volume limit can be approached in a 
meaningful way.

\subsection{Periodic Versus C-Periodic Spatial Boundary Conditions}

Most numerical studies of lattice gauge theories use periodic spatial boundary 
conditions,
\begin{equation}
A_\mu(\vec x + L_i \vec e_i,x_4) = A_\mu(\vec x,x_4).
\end{equation}
Here $L_i$ is the size of a 3-dimensional torus in the $i$-direction and 
$\vec e_i$ is the unit-vector pointing in that direction. However, periodic 
boundary conditions are not the most useful ones for studying a single Polyakov
loop. As we have seen, a Polyakov loop represents a single static quark that 
transforms non-trivially under the center of the gauge group. Indeed, the test 
quark carries a non-zero $\Z(N)_c$ charge. For topological reasons, a single 
charged particle cannot exist in a periodic volume. The flux emanating from the
charge cannot escape to infinity and consequently the Gauss law demands that it
must end in a compensating anti-charge \cite{Hil83}. As a whole, a periodic 
system is always neutral. 

Here we consider periodic boundary conditions in the $x$ and $y$-directions and
$C$-periodic boundary conditions in the $z$-direction \cite{Pol91,Kro91}. When 
a $C$-periodic field is shifted by $L_z$, it is replaced by its charge 
conjugate, i.e. for $C$-periodic gluons
\begin{equation}
A_\mu(\vec x + L_z \vec e_z,t) = \ ^C\!A_\mu(\vec x,t) = A_\mu(\vec x,t)^*.
\end{equation}
In a $C$-periodic volume, a single quark can exist, because now its center 
electric flux can escape to its charge conjugate partner on the other side of 
the boundary. Note that the system is still translationally invariant. The 
allowed gauge transformations, as well as the Polyakov loop, also satisfy 
$C$-periodicity
\begin{equation}
g(\vec x + L_z \vec e_z,t) = g(\vec x,t)^*, \
\Phi(\vec x + L_z \vec e_z) = \Phi(\vec x)^*.
\end{equation}
As a consequence of the boundary conditions, the $\Z(3)_c$ center symmetry is 
now explicitly broken \cite{Wie92}. If one again considers a transformation 
$g(\vec x,t + \beta) = z g(\vec x,t)$, which is periodic, up to a center 
element $z \in \Z(3)_c$, in the Euclidean time direction, one finds
\begin{equation}
g(\vec x,t)^* = g(\vec x,t + \beta)^* z = g(\vec x + L_z \vec e_z,t + \beta) z
= g(\vec x + L_z \vec e_z,t) z^2 = g(\vec x,t)^* z^2.
\end{equation}
Consistency requires $z^2 = 1$ and hence $z = 1$ (because $z \in \Z(3)_c$). Of
course, in the infinite volume limit, the explicit $\Z(3)_c$ symmetry breaking
due to the spatial boundary conditions disappears. With $C$-periodic boundary
conditions, $\langle \Phi \rangle$ is always non-zero in a finite volume. In 
the confined phase $\langle \Phi \rangle$ goes to zero in the infinite volume 
limit, while it remains finite in the high-temperature deconfined phase. 
$C$-periodic boundary conditions are well-suited for studying the free energy 
of single quarks, while with periodic boundary conditions a single quark cannot
even exist.

\subsection{Static Quarks in $C$-periodic Cylinders below and above $T_c$}

First, we consider the system in the confined phase at temperatures $T < T_c$ 
in a cylindrical volume with cross section $A = L_x L_y$ and length 
$L_z \gg L_x,L_y$. Diagrammatically, the partition function is
\begin{equation}
Z = \begin{picture}(90,15)
\put(0,-4){\line(1,0){90}} \put(0,11){\line(1,0){90}}
\put(0,-4){\line(0,1){15}} \put(90,-4){\line(0,1){15}} \put(43,0){$c$}
\end{picture} \ = \exp(- \beta f_c A L_z),
\end{equation}
where $f_c$ is the temperature-dependent free energy density in the confined
phase. The expectation value of the Polyakov loop (times $Z$), on the other
hand, is given by
\begin{equation}
Z \langle \Phi \rangle = \begin{picture}(90,15)
\put(0,-4){\line(1,0){90}} \put(0,11){\line(1,0){90}}
\put(0,-4){\line(0,1){15}} \put(90,-4){\line(0,1){15}} \put(43,0){$c$}
\put(0,3.5){\line(1,0){38}} \put(52,3.5){\line(1,0){38}}
\end{picture} \ = \exp(- \beta f_c A L_z) \Sigma_0 \exp(- \beta \sigma L_z),
\end{equation}
where $\sigma$ is the string tension, which is again temperature-dependent. The
confining string (denoted by the additional line in the diagram) connects the 
static quark with its anti-quark partner on the other side of the $C$-periodic 
boundary. Hence, the free energy of the quark is given by
\begin{equation}
F = - \frac{1}{\beta} \log \Sigma_0 + \sigma L_z.
\end{equation}
The free energy diverges as $L_z \rightarrow \infty$, indicating that the quark
is confined.

Now let us consider the deconfined phase at temperatures $T > T_c$, where three
distinct deconfined phases coexist. In a cylindrical volume, a typical 
configuration consists of several bulk phases, aligned along the $z$-direction,
separated by deconfined-deconfined domain walls. The domain walls cost free 
energy $F$ proportional to their area $A$, such that their tension is given by 
$\alpha_{dd} = F/A$. What matters in the following is the cylindrical shape, 
not the magnitude of the volume. In fact, the cylinders can be of macroscopic 
size. The expectation value of the Polyakov loop in a cylindrical volume can be
calculated from a dilute gas of domain walls \cite{Hol97}. The domain wall 
expansion of the partition function can be viewed as
\begin{equation}
Z = \begin{picture}(90,15)
\put(0,-4){\line(1,0){90}} \put(0,11){\line(1,0){90}}
\put(0,-4){\line(0,1){15}} \put(90,-4){\line(0,1){15}} \put(40,0){$d^{(1)}$}
\end{picture} \ + \
\begin{picture}(90,15)
\put(0,-4){\line(1,0){90}} \put(0,11){\line(1,0){90}}
\put(0,-4){\line(0,1){15}} \put(90,-4){\line(0,1){15}} 
\put(45,-4){\line(0,1){15}} \put(18,0){$d^{(2)}$} \put(63,0){$d^{(3)}$} 
\end{picture} \ + \
\begin{picture}(90,15)
\put(0,-4){\line(1,0){90}} \put(0,11){\line(1,0){90}}
\put(0,-4){\line(0,1){15}} \put(90,-4){\line(0,1){15}}
\put(45,-4){\line(0,1){15}} \put(18,0){$d^{(3)}$} \put(63,0){$d^{(2)}$}
\end{picture} \ + ...
\end{equation}
The first term has no domain walls and thus the whole cylinder is filled with
deconfined phase $d^{(1)}$ only. An entire volume filled with either phase 
$d^{(2)}$ or $d^{(3)}$ would not satisfy the boundary conditions. The second 
and third terms have one domain wall separating phases $d^{(2)}$ and $d^{(3)}$.
Here, $C$-periodic boundary conditions exclude phase $d^{(1)}$. The sum of the 
diagrammatic terms is given by
\begin{eqnarray}
Z&=&\exp(- \beta f_d A L_z) \nonumber \\
&+& 2 \int_0^{L_z} dz \ \exp(- \beta f_d A z)
\gamma \exp(- \beta \alpha_{dd} A) \exp(- \beta f_d A (L_z - z)) + ... 
\nonumber \\
&=&\exp(- \beta f_d A L_z)
[1 + 2 \gamma \exp(- \beta \alpha_{dd} A) L_z + ...].
\end{eqnarray}
The first term is the Boltzmann weight of deconfined phase of volume $A L_z$
with a free energy density $f_d$. The second term contains two of these bulk
Boltzmann factors separated by a domain wall contribution 
$\gamma \exp(- \beta \alpha_{dd} A)$, where $\gamma$ is a factor resulting from
capillary wave fluctuations of the domain wall. In the above expression, we 
have integrated over all possible locations $z$ of the domain wall. It is 
straightforward to sum the domain wall expansion to all orders, giving
\begin{equation}
Z = \exp(- \beta f_d A L_z + 2 \gamma \exp(- \beta \alpha_{dd} A) L_z).
\end{equation}
In exactly the same way one obtains
\begin{equation}
\langle \Phi \rangle = \Phi_0 \exp(- 3 \gamma \exp(- \beta \alpha_{dd} A) L_z).
\end{equation}
The free energy of a static quark in a $C$-periodic cylinder is therefore given
by
\begin{equation}
F = - \frac{1}{\beta} \log \Phi_0 + 
\frac{3 \gamma}{\beta} \exp(- \beta \alpha_{dd} A) L_z.
\end{equation}
This result is counter intuitive. Although we are in the deconfined phase, the
quark's free energy diverges in the limit $L_z \rightarrow \infty$, as long as
the cross section $A$ of the cylinder remains fixed. This is the behavior one
typically associates with confinement. In fact, 
\begin{equation}
\sigma' = \frac{3 \gamma}{\beta} \exp(- \beta \alpha_{dd} A)
\end{equation}
plays the role of a ``string tension'', even though there is no physical string
that connects the quark with its anti-quark partner on the other side of the 
$C$-periodic boundary. ``Confinement'' in $C$-periodic cylinders arises because
disorder due to many differently oriented deconfined phases destroys the 
correlations of center electric flux between quark and anti-quark.

Of course, this paradoxical confinement mechanism is due to the cylindrical
geometry and the specific boundary conditions. Had we chosen $C$-periodic 
boundary conditions in all directions, the deconfined phases $d^{(2)}$ and 
$d^{(3)}$ would be exponentially suppressed, so that the entire volume would be
filled with phase $d^{(1)}$ only. In that case, the free energy of a static 
quark is $F = - (1/\beta) \log \Phi_0$, which does not diverge in the infinite 
volume limit, as expected. Had we worked in a cubic volume, a typical 
configuration would have no domain walls, the whole volume would be filled with
deconfined phase $d^{(1)}$ and again the free energy of a static quark would be
$F = - (1/\beta) \log \Phi_0$. Finally, note that, in a cylindrical volume, 
even the static Coulomb potential is linearly rising with a ``string tension''
$e^2/A$. Due to Debye screening, this trivial confinement effect is absent
in the gluon plasma.

The observed paradoxical confinement implies that deconfined-deconfined domain
walls are more than just Euclidean field configurations. In fact, they can
lead to a divergence of the free energy of a static quark and thus they have 
physically observable consequences even in Minkowski space-time. Of course, the
issue is rather academic. First of all, the existence of three distinct 
deconfined phases relies on the $\Z(3)_c$ symmetry, which only exists in a pure
gluon system --- not in the real world with light dynamical quarks. Second, the
effect is due to the cylindrical geometry and our choice of boundary 
conditions. Even though it is not very realistic, this set-up describes a 
perfectly well-defined Gedanken experiment, demonstrating the physical reality 
of deconfined-deconfined domain walls.

\section{Conclusions}

Thanks to the efforts of numerous people, the role of the center symmetry for
the confinement and deconfinement of static quarks is by now very well 
understood and is more or less a closed subject. In this article we have 
concentrated on those aspects of the problem that can be easily explained using
simple analytic calculations in continuum field theory. However, we again want 
to emphasize the very important role that lattice field theory, in particular,
numerical simulation of $SU(2)$ and $SU(3)$ Yang-Mills theory has played in 
understanding the dynamical role of the center symmetry. It is fair to say 
that lattice gauge theory has led to a reliable high-precision numerical 
solution of Yang-Mills theories. Due to severe numerical problems in dealing 
with fermionic fields, this can presently not be said about full QCD including 
dynamical quarks. Still, continuous progress is being made in applying lattice 
methods and it is to be expected that an accurate numerical solution of QCD 
will eventually be obtained. In contrast to Yang-Mills theory, the center
symmetry plays a minor role in full QCD because it is explicitly broken in the 
presence of quarks. However, along the way, other theories --- for example, 
supersymmetric Yang-Mills theories --- can be solved with lattice field theory 
methods as well. At that point the center symmetry may become an interesting 
research topic again. In any case, we hope that we have convinced the reader 
that --- thanks to the center symmetry --- a world without quarks would not be 
quite as boring as one might have naively expected.

\section*{Acknowledgments}

We like to thank M. Shifman for inviting us to contribute to this volume. We 
are also grateful to our collaborators M. Alford, A. Campos, 
S. Chandrasekharan, J. Cox, B. Grossmann, A. Kronfeld, M. Laursen, L. Polley 
and T. Trappenberg with whom we have explored some of the physics described in 
this article during the past ten years. We are also indebted to A. Smilga for 
numerous discussions on the nature and the physical relevance of the center 
symmetry. U.-J. W. likes to thank the theory group of Erlangen University, 
where this article was completed, and especially F. Lenz for hospitality. K. H.
acknowledges the support of the Schweizerischer Nationalfond. U.-J. W. is
supported in part by funds provided by the U.S. Department of Energy (D.O.E.)
under cooperative research agreements DE-FC02-94ER40818 and also by the 
A. P. Sloan foundation.

\end{document}